\documentstyle[prl,aps,twocolumn,epsf]{revtex}
\newcommand*{\be}{\begin{equation}} 
\newcommand*{\ee}{\end{equation}}

\begin{document}

\title{ \bf Large-$N$ $\beta$-function for superconducting films in a 
magnetic field}
 
\author{A.P.C. Malbouisson${}^{(a)}$} 

\address{ (a) {\it CBPF/MCT - Rua Dr. Xavier Sigaud 150, Urca,} \\   
{\it Rio de Janeiro CEP 22290-180-RJ, Brazil. E-mail: adolfo@lafex.cbpf.br}}

\vspace{0.7cm}

\maketitle

\begin{abstract}
{\bf Abstract} By considering the large-N Ginzburg-Landau model, compactified in one of the spatial 
dimensions,  we determine the 
beta-function and find an infrared stable fixed point for a superconducting film 
in presence of an external magnetic field, for dimensions $4<D<6$. We 
find that this fixed point is independent of the 
film thickness ($L$). For $D\leq 4$, severe infrared divergences 
appearing under the form of 
 divergence of the integral in the quantity $a(D)$ defined after Eq.(\ref{sigma2}),
forbid an analogous analysis. 
For the physical dimension $D=3$ we conclude that if a transition 
exists, it should not be a second order one.\\
PACS number(s): 74.20.-z, 05.10Cc, 11.25.Hf

% 
%\vspace{0.3cm}
%
\noindent
PACS number(s): 74.20.-z, 05.10Cc, 11.25.Hf
\end{abstract}

It is usually assumed that it is a good approximation to neglect magnetic thermal 
fluctuations in the Ginzburg-Landau (GL) model, when applied to study high temperature superconductors. This is due to the fact that these materials have a very large Ginzburg 
parameter, typically $\kappa\sim 100$.
This problem has been investigated by a number of authors, both in its single 
component and in its $N$- component versions. A large account on the 
state of the subject and related topics 
can be found for instance in 
refs. \cite{Afleck,Lawrie,Lawrie1,Brezin,Moore,Radz,Flavio} 
and other references therein. In particular in ref.\cite{Radz} a large-$N$ theory of 
a second order transition for arbitrary dimension $D$ is presented and the fixed point 
effective free energy describing the transition is found. This author claims that 
it is possible that in the physical situation of $N=1$, a mechanism of reduction 
of the lower critical domension could allow a continuous transition in $D=3$. 
Here we investigate a confined version of the model studied in ref.\cite{Radz}. 
We investigate the large-$N$ limit of the Ginzburg-Landau model, 
the system being confined between two parallel 
planes a distance $L$ apart from one another. Studies on confined field theory 
have been done in the literature since a long time ago. In particular, an analysis 
of the renormalization group in finite size geometries can be found in ref.\cite{Zinn}. 
This study is performed using a modified Matsubara formalism to take into account 
boundary effects on scaling laws. 

In this note we use an extended  
 compactification formalism  in the framework of the effective potential, introduced in 
 a recent publication \cite{Ademir}, where 
 the Euclidean massive $(\lambda\varphi^{4})_{D}$ vector $N$-component model 
has been considered, and  a study of the system 
 subject to confinement on a $d$-dimensional subspace, $d\leq D$, has been performed. This 
allowed to generalize to a $d$-dimensional   
subspace, previous results in the effective potential framework  
 for finite temperature and spacial boundaries. In particular it has been shown how a 
 compactification mechanism applies to the study of spatially confined systems, 
 for finite or zero-temperature field theories. This mechanism 
generalizes and unifies  
results from a recent work 
on the behaviour of field theories in presence of spatial planar boundaries
\cite{JMario,FAdolfo}, and 
previous 
 results in the literature for finite temperature field theory, as for instance in \cite{Gino1}. 
 In the present work  
our aim is to show  how the above mentioned formalism can be used  to study in a field theoretical framework, effects associated to 
 spacial confinement, for a model that allows a
non-perturbative approach and is supposed to have a direct physical interpretation.
 We consider the vector $N$-component 
Ginzburg-Landau model in presence of an external magnetic field 
at leading order in $\frac{1}{N}$, the system 
being  submited to the constraint of being confined between two parallel planes 
a distance $L$ apart from one another.  For 
$\kappa\gg 1$ the Hamiltonian density of the 
GL model in an external magnetic field can be written in the form, 
\begin{equation}
\label{GL}
{\cal H}=|(\nabla-ie{\bf A})\phi|^2+m^{2}|\phi|^{2}+\frac{u}{2}|\phi|^4,
\end{equation}
where $\nabla\times{\bf A}={\bf H}$ and $m^2=a(T-T_c)$ with 
$a>0$. This model Hamiltonian describes superconductors in the 
extreme type II limit. In the following we assume that the external magnetic field is 
parallel to the $z$ axis and that the gauge ${\bf A}=(0,xH,0)$ has 
been chosen. We will consider the model (\ref{GL}) with $N$ complex 
components and take the large $N$ limit at $Nu$ fixed.
If we consider the system in unlimited space, the field $\phi$ should be 
written in terms of the well known Landau level basis, 
\begin{equation}
\phi({\bf r})=\sum_{l=0}^{\infty}\int\frac{dp_y}{2\pi}\int\frac{d^{D-2}p}{(2\pi)^{D-2}}
\hat{\phi}_{l,p_y,{\bf p}}\chi_{l,p_y,{\bf p}}({\bf r}),
\label{landau}
\end{equation}
where $\chi_{l,p_y,p_z}({\bf r})$ are the Landau level eingenfunctions given 
by
\begin{eqnarray}
\chi_{n,p_y,p_z}({\bf r})&=&\frac{1}{\sqrt{2^l}l!}\left(\frac{\omega}{
\pi}\right)^{1/4}e^{i({\bf p}\cdot {\bf u}+p_y y)}e^{-\omega(x-p_y/\omega)^2/2}\times  \nonumber \\
& & \times H_l\left(\sqrt{\omega}x-\frac{p_y}{\sqrt{\omega}}\right), \nonumber \\
\end{eqnarray}
with energy eigenvalues 
$E_{l}(|{\bf p}|)=|{\bf p}|^{2}+(2l+1)\omega+m^{2}$
and where $H_l$ are the Hermite polynomials. In the above equations ${\bf p}$ and 
${\bf u}$ are $(D-2)$-dimensional vectors.

Now, let us consider
the system confined between two parallel planes, normal to the $z$-axis, a
distance $L$ apart from one another and use Cartesian coordinates ${\bf r} 
=(z,{\bf z})$, where ${\bf z}$ is a $(D-3)$-dimensional vector, with
corresponding momenta ${\bf k}=(k_{z},{\bf q})$, ${\bf q}$ being a $(D-3)$ 
-dimensional vector in momenta space. In this case, the model is supposed 
to describe a superconducting material in the form of a film.
Under these conditions the generating functional of the correlation functions is written as,
\begin{equation}
{\cal Z}=\int {\cal D}\phi^{\ast} {\cal D}\phi exp\left(-\int_{0}^{L} dz \int d^{D-3} {\bf z} \;{\cal H}(|\phi|, |\nabla \phi|\right),
\label{Z}
\end{equation}
with the field $\phi(z,{\bf z})$ satisfying the condition of confinement along
the $z$-axis, $\varphi(z=0, {\bf z})\;=\;\varphi(z=L, {\bf z})$. 
Then the field representation (\ref{landau}) should be modified and 
 have a mixed series-integral Fourier expansion of the 
form,
\begin{eqnarray}
\phi(z,{\bf z})&=&\sum_{l=0}^{\infty}\sum_{n=-\infty}^{\infty} c_{n}\int\frac{dp_y}{2\pi}\int d^{D-3} {\bf q} \;b({\bf q})\times  \nonumber \\ & &  \times e^{-i\omega_{n} x\;-i{\bf q}\cdot {\bf z}} \tilde{\varphi}_{l}(\omega_{n}, {\bf q}), \nonumber \\
\label{Fourier}
\end{eqnarray}
where $\omega_{n}=2\pi n/L$, the label $l$ 
refers to the Landau levels, and the coefficients $c_{n}$ and $b({\bf q})$   
correspond respectivelly to the Fourier series representation over $z$ and 
  to the Fourier integral representation over the $D-3$-dimensional 
${\bf z}$-space.
The above conditions of confinement of the $z$-dependence of the
field to a segment of length $L$, allow us to 
proceed with respect to the $z$-coordinate, 
in a manner analogous as it is done in the imaginary-time 
Matsubara formalism in field theory. The Feynman rules should be modified 
following the prescription,
\begin{equation}
\int \frac{dk_{z}}{2\pi }\rightarrow \frac{1}{L}\sum_{n=-\infty }^{+\infty
}\;,\;\;\;\;\;\;k_{z}\rightarrow \frac{2n\pi }{L}\equiv \omega _{n}.
\label{Matsubara}
\end{equation}
We emphasize that here we are considering an
Euclidean field theory in $D$ {\it purely} spatial dimensions, we are {\it not} 
working in the framework of finite temperature field theory.   
Temperature is introduced in the mass term of the Hamiltonian by means of the 
usual Ginzburg-Landau prescription. In the following we consider only the lowest Landau 
level $l=0$.
Thus, 
limiting ourselves to an analysis at the lowest Landau level (LLL), we obtain that
the effective $|\phi|^4$ interaction is given in momentum space 
and at the critical point by
\begin{equation}
\label{interac}
U({\bf p})=
\frac{u}{1+Nu\omega
e^{-\frac{1}{2\omega}(p_x^2+p_y^2)}\Sigma(p)},
\label{g}
\end{equation}
where $\Sigma(p)$, the single one-loop buble is given by,
\begin{eqnarray}
\Sigma(p)&=&\frac{1}{L}\sum_{n=-\infty}^{\infty}\int_{0}^{1}dx\int\frac{d^{D-3}k}{(2\pi)^{D-3}}\times  \nonumber \\
& & \times \frac{1}{\left[k^{2}+\omega_{n}^{2}+p^{2}x(1-x)\right]^{2}}. \nonumber \\
\label{sigma1}
\end{eqnarray}
The sum over $n$ and the integral over $k$ can be treated using the formalism developed in 
\cite{Ademir}. The starting point is an expression of the form, 
\begin{equation}
U=\sum_{n=-\infty }^{+\infty }\int \frac{d^{D-1}k}{(an^{2}+c^{2}+ 
{\bf k}^{2})^{s}},  \label{potefet1}
\end{equation}
which, using a well-known dimensional regularization formula \cite{Zinn}, can be written 
in the form,
\begin{equation}
U=f(D,s)g^{s}Z_{1}^{c^{2}}(s-\frac{D-1}{2};a),  \label{potefet2}
\end{equation}
where $f(D,s)$ is a function proportional to $\Gamma (s-\frac{D}{2})$ and $ Z_{1}^{c^{2}}(s-\frac{D}{2};a)$ is one of the Epstein-Hurwitz $zeta$
-functions defined by, 
\begin{equation}
Z_{K}^{c^{2}}(u;\{a_{i}\})=\sum_{n_{1},...,n_{K}=-\infty }^{+\infty
}(a_{1}n_{1}^{2}+...+a_{K}n_{K}^{2}+c^{2})^{-u},  \label{zeta}
\end{equation}
valid for $Re(u)>\;K/2$ (in our case $Re(s)>\;D/2$). The Epstein-Hurwitz $ 
zeta$-function can be extended to the whole complex $s$-plane and we obtain,
after some manipulations \cite{Ademir}, \begin{eqnarray}
U&=& h(D,s)\left[2^{-(\frac{D}{2}-s+2)}\Gamma(s-\frac{D}{2})(m/\mu)^{D-2s}+\right.\nonumber \\
& &  \left.+\sum_{n=1}^{\infty
}(\frac{m}{\mu ^{2}nL})^{\frac{D}{2}-s}K_{\frac{D}{2}-s}(mnL)\right],\nonumber \\
\label{potefet3}
\end{eqnarray}
where 
\begin{equation}
h(d,s)=\frac{2}{2^{D/2-s-1}\pi ^{D/2-2s}}\frac{1}{\Gamma
(s)}
\label{h}
\end{equation}
 and $K_{\nu }$ are the Bessel functions of the third kind.\\
Applying formula (\ref{potefet3})  to Eq.(\ref{sigma1}) the result is,
\begin{eqnarray}
\Sigma(p)&=& (2\pi)^{\frac{10-D}{2}}\left[2^{\frac{-D}{2}}a(D)\Gamma\left(\frac{6-D}{2}\right)(p^{2})^{\frac{D-6}{2}}+\right.\nonumber \\
& & \left.+\int_{0}^{1}dx\sum_{n=1}^{\infty }\left(\frac{\sqrt{p^{2}x(1-x)}}{nL}\right)^{\frac{D-6}{2}} \times \right. \nonumber \\
& & \left. K_{\frac{D-6}{2}
}\left(nL \sqrt{p^{2}x(1-x)}\right),\right]\nonumber \\
\label{sigma2}
\end{eqnarray}
where $a(D)=\int_{0}^{1}dx[x(1-x)]^{\frac{D-6}{2}}$. 

If an infrared stable fixed point exists, it would be possible to determine it by 
an study of the infrared behaviour of the 
beta-function, $i.e$, in the neighbourhood of $|{\bf p}|=0$. Therefore 
we should investigate the above equations for $|{\bf p}|\approx 0$. We start from 
(\ref{sigma2}), using an asymptotic formula for small values of the argument of Bessel
functions, 
\begin{equation}
K_{\nu}(z)\approx \frac{1}{2}\Gamma(\nu)\left(\frac{z}{2}\right)^{-\nu}\;\;\;(z\sim 0\;;\;\;\;Re(\nu)>0),
\label{K}
\end{equation}
which allows after some straighforward manipulations, to write Eq.(\ref{sigma2}) 
for $|{\bf p}|\approx 0$ in the form,
\begin{eqnarray}
\Sigma(p)&\approx & (2\pi)^{\frac{10-D}{2}}\left[2^{\frac{D-4}{2}}a(D)\Gamma\left(\frac{6-D}{2}\right)|{\bf p}|^{D-6}+\right.\nonumber \\
& & \left.+L^{6-D}2^{\frac{D-10}{2}}\Gamma\left(\frac{D-6}{2}\right)\zeta(D-6)\right],\nonumber \\
\label{sigma3}
\end{eqnarray}
valid for {\it odd} dimensions $D>7$, due to the poles of the $\Gamma$ and $\zeta$ functions. 
We can obtain an expression for smaller values of $D$ performing an analytic continuation 
of the Riemann $zeta$-function $\zeta(D-7)$ 
by means of the reflexion property of $zeta$-functions, 
\begin{equation}
\zeta (z)=\frac{1}{\Gamma (z/2)}\Gamma (\frac{1-z}{2})\pi ^{z-\frac{1}{2} 
}\zeta (1-z),
  \label{extensao}
\end{equation}
which gives,
\begin{equation}
\zeta (D-6)=\frac{1}{\Gamma (\frac{D-6}{2})}\Gamma (\frac{7-D}{2})\pi ^{\frac{2D-15}{2} 
}\zeta (7-D).
  \label{extensao1}
\end{equation}
Inserting Eq.(\ref{extensao1}) in Eq.(\ref{sigma3}), and remembering the definition of 
$a(D)$ after Eq.(\ref{sigma2}), 
we obtain an expression valid 
for $4<D<6$,
\begin{equation}
\Sigma(p)\approx  A(D)|{\bf p}|^{D-6}+B(D,L),
\label{sigma4}
\end{equation}
where,
\begin{equation}
B(D,L)=\pi^{\frac{D-3}{2}}L^{6-D}\Gamma\left(\frac{7-D}{2}\right)\pi ^{\frac{2D-13}{2}}\zeta(7-D)
\label{B}
\end{equation}
and 
\begin{equation}
A(D)=(2\pi)^{\frac{10-D}{2}}2^{\frac{4-D}{2}}\Gamma\left(\frac{6-D}{2}\right)a(D).
\label{A}
\end{equation}
Inserting (\ref{B}) and (\ref{A}) in Eq.(\ref{g}) we have,
\begin{equation}
g(|{\bf p}|\approx 0)\approx  \frac{u}{1+Nu\omega
\left[A(D)|{\bf p}|^{D-6}+B(D,L)\right]}.
\label{g3}
\end{equation}
Let us take as a running scale $|{\bf p}|$, and define 
the dimensionless coupling 
\begin{equation}
g=U_\sigma(p_x=0,p_y=0,{\bf p})\omega |{\bf p}|^{D-6},
\label{g1}
\end{equation}
where we remember that ${\bf p}$ is a $D-3$-dimensional vector. 
Then we obtain straightforwardly the $beta$-function for $|{\bf p}|\approx 0$,
\begin{equation}
\beta(g)=|{\bf p}|\frac {\partial g}{\partial |{\bf p}|}\approx (6-D)\left[-g+NA(D)g^{2}\right]
\label{beta}
\end{equation}
from which we get the infrared stable fixed point,
\begin{equation}
g_{*}(D)=\frac{1}{NA(D)}.
\label{gstar}
\end{equation}

Previous renormalization group calculations for materials in bulk form 
in ref.\cite{Brezin} indicated a 
first order transition for $4<D<6$. This result has been obtained using an 
$\epsilon=6-D$ expansion. The same conclusion of a first order transition is 
obtained in Ref.\cite{Afleck} with a large-N calculation. A large-N analysis 
and a functional renormalization group study performed 
in Refs.\cite{Moore,Radz,Moore1} leads to an opposite result to that of Ref.\cite{Afleck},
 concluding for a second order transition in dimensions $4<D<6$. The same 
conclusion is obtained in Ref.\cite{Flavio}.  
The authors in Ref.\cite{Moore} claim 
moreover that the inclusion of fluctuations do not alter significantly the main characteristic  
of the system, that is, the existence of 
a continuous transition into a spatially homogeneous condensate. 
Therefore for $4<D<6$ our result for a film-like 
material (finite $L$) is the existence of an infrared stable fixed point, 
in agreement with those obtained for the material in bulk form 
($L=\infty$) in   
Refs.\cite{Moore,Radz,Flavio}. Moreover the fixed point 
is independent of the film thickness $L$. This could be interpreted as 
pointing to a second order transition.  However this should be taken as an {\it indication} 
not as a demonstration of the existence of a continuous transition.  
As already discussed in \cite{Moore,Moore1}, even if infrared fixed points exist none of them 
can be completely atractive, due to the large space of couplings. 
In this case the existence of an infrared 
fixed point as found in this paper, does not assure the existence of a second order transition. 
For $D\leq 4$, severe infrared divergences 
appear under the form of 
 divergence of the integral in the quantity $a(D)$ defined after Eq.(\ref{sigma2}).
We conclude that, for materials under the form of films, as 
is also the case for materials in bulk form, if there exists 
a phase transition for $D\leq 4$, in particular in $D=3$, it should not be a second order one.\\

This paper has been partially supported by the Brazilian agencies CNPq (Brazilian 
National Research Council) and
FAPERJ (Foundation for the support of research in the state of Rio de Janeiro).


\begin{references}
\bibitem{Afleck}  I. Affleck and E. Br\'{e}zin, Nucl. Phys. {\bf 257}, 451
(1985).

\bibitem{Lawrie}  I. D. Lawrie, Phys. Rev. B {\bf 50}, 9456 (1994).

\bibitem{Lawrie1}  I. D. Lawrie, Phys. Rev. Lett. {\bf 79}, 131 (1997).

\bibitem{Brezin}  E. Br\'{e}zin, D. R. Nelson, and A. Thiaville, Phys. Rev.
B {\bf 31}, 7124 (1985).

\bibitem {Moore} M.A. Moore, T.J. Newman, A.J.Bray, S.-K. Chin, Phys. Rev. B,{\bf 58}, 936 (1998)

\bibitem{Radz} L. Radzihovsky, Phy. Rev. Lett. {\bf 74}, 4722 (1995); 
{\it ibid.} {\bf 76}, 4451 (1996); I. F. Herbut and Z. Tesanovic, 
{\it ibid.} {\bf 76}, 4450 (1996)

 \bibitem{Flavio}  C. de Calan, A. P. C. Malbouisson, and F. S. Nogueira,
Phys. Rev. B, {\bf 64}, 212502 (2001).

\bibitem{Zinn}  J. Zinn-Justin, {\it Quantum Field Theory and Critical
Phenomena} (Clarendon Press, Oxford, 1996).

\bibitem{Ademir}  A. P. C. Malbouisson, J. M. C. Malbouisson, A. E.
Santana, Nucl. Phys. B, {\bf 631}, 83 (2002)

\bibitem{JMario}  A. P. C. Malbouisson and J. M. C. Malbouisson, J. Phys. A:
Math. Gen. {\bf 35}, 2263 (2002).

\bibitem{FAdolfo}  L. Da Rold, C. D. Fosco, and A. P. C. Malbouisson, Nucl.
Phys. B {\bf 624}, 485 (2002).

\bibitem{Gino1}  G. N. J. A\~{n}a\~{n}os, A. P. C. Malbouisson, and N. F.
Svaiter, Nucl. Phys. B {\bf 547}, 221 (1999).

\bibitem{Moore1} T.J. Newman, M.A. Moore, Phys. rev. B, {\bf 54}, 6661 (1996)


\end{references}
\end{document}